\documentclass[12pt, titlepage]{article}%
\usepackage{eurosym}
\usepackage{amsmath,amssymb,amsthm,amsfonts}
\usepackage[mathscr]{eucal}
\usepackage{graphicx}
\usepackage{epstopdf}
\usepackage{subcaption}
\usepackage{appendix}
\usepackage[comma,authoryear]{natbib}
\usepackage{amsmath}
\usepackage{amsfonts}
\usepackage{amssymb}
\usepackage{multirow}
\usepackage{rotating}
\usepackage{xcolor}
\usepackage{lscape}
\usepackage{float}
\usepackage{appendix}
\usepackage{authblk}
\usepackage{authblk}%
\providecommand{\U}[1]{\protect\rule{.1in}{.1in}}
\numberwithin{equation}{section} 

\newcommand{\ba}{\begin{eqnarray}}
\newcommand{\ea}{\end{eqnarray}}

\title{Non-maturing deposits modelling in a Ornstein-Uhlenbeck framework}
\author[1]{Marina Marena}
\author[2]{Andrea Romeo}
\author[3]{Patrizia Semeraro}
\affil[1]{
Department of Economics and Statistics,  Universit\`a di Torino.}
\affil[2]{
European Banking Authority.}
\affil[2]{Department of Mathematical Sciences G. Lagrange, Politecnico di Torino.}
\date{}                     
\begin{document}
%
\maketitle

\begin{abstract}
This paper builds a multivariate L\'evy-driven Ornstein-Uhlenbeck process for
the management of non-maturing deposits, that are a major source of funding
for banks. The contribution of the paper is both theoretical and operational.
On the theoretical side, the novelty of this model is to include three
independent sources of randomness in a L\'evy framework: market interest rates, deposit rates and
deposit volumes.
The choice of a L\'evy background driving process allows us to model rare but
severe events. On the operational side, we propose a procedure to include severe volume outflows with positive
	probability in future scenarios simulation, explaining its implementation with an illustrative example using Italian banking sector data.

\noindent\textbf{Journal of Economic Literature Classification}: G12, G13


\end{abstract}



\section*{Introduction}

Non-maturing deposits {(NMDs)} are a major source of funding for financial
institutions. Many definitions for {NMDs} can be provided.
According to the Basel Committee on Banking Supervision (\cite{BCBS2016IRRBB}%
), {NMDs} are liabilities of the banks where there is no
contractual maturity and consequently where depositors are free to withdraw
them at any time, partially or entirely. On top of the possibility for
depositors to withdraw their money at any time, another intrinsic
characteristic of {NMDs} is the faculty for banks to reprice
these products at any time. Several types of products issued by banks can fall
under this definition, including (in a non-exhaustive list): non-maturing
liability checking accounts, savings accounts and short time deposits. {NMDs are}
important for banks, due to their stability and cheapness on the
liability side. These characteristics become increasingly important in periods
when market turmoil threatens to preclude other sources of funding.
Considering that {NMDs} entail these two features, managing
them is anything but a trivial task for banks. In addition, during the few
past years (mostly after the financial crisis), a number of credit institutions
experienced or almost experienced unexpected shortages --- the so-called "bank
run". A number of factors can produce this phenomenon, mainly involving a lack
of confidence in the creditworthiness and accountability of the bank.
Consequently there is a clear need for a modelling approach for {NMDs} which is appropriate, effective and prudent. Although the topic has
not been extensively investigated in the literature, different approaches have
been proposed to tackle this issue, e.g. \cite{jarrow1998arbitrage},
\cite{kalkbrener2004risk}, \cite{blochlinger2015identifying} and
\cite{castagna2017} to cite just some of them.

This work has {both} theoretical and operational contributions. The theoretical
contribution of the present work is to introduce a multivariate
Ornstein-Uhlenbeck (OU) process to model the interactions among market interest
rates, deposit rates and deposit volumes. The novelty of our approach is to
explicitly model three independent sources of randomness, each one affecting
one or more of the one-dimensional marginal driving processes within our
multivariate framework. These three explicit sources of randomness are assumed
to be the implicit options held by the bank and the customers, alongside
market interest rates. The idea is that, in order to adequately capture the
possibility for a single bank of facing stressed situations (irrespective of
the general economic environment), both the deposit rates and the volumes of
deposits should depend on idiosyncratic components specific to the bank and
its customers. Given the economic relationships among the modelled variables, a
triangular specification of the multivariate background driving process is
assumed. Driving processes belonging to the L\'{e}vy class are considered { to be fit for our purpose}.
This class of processes are in our view particularly suitable for the purpose
of {NMDs} modelling, for their ability to accommodate
skewness and {excess} kurtosis (\cite{Ba} and \cite{masuda2004multidimensional}).
Indeed, both skewness and {excess} kurtosis play a key role in modelling rare
events that can entail a significant risk in terms of liquidity management
for a bank.

 {The estimation procedure is based on maximum likelihood (ML) methods applied in our multidimensional setting.}
The operational contribution is to introduce an \textit{ad-hoc} procedure to simulate future
cash flows in stressed scenarios which encompass rare but severe volume outflows. This procedure can be used for liquidity risk management purposes.

In what follows, Section \ref{sect:levyOU} presents the stochastic model and
Section \ref{sect:appl} describes the estimation procedure. Then, in Section
\ref{sect:fb}, the parameters of the model are estimated. The application to
liquidity risk management for banks is in Section \ref{sec:lrm}. Section
\ref{conclusion} concludes.

\section{The model: L\'evy-driven OU process \label{sect:levyOU}}

Let $\boldsymbol{L}(t)$ be a L\'{e}vy process on $\mathbb{R}^{n}$ starting
from the origin with independent components. Let $\boldsymbol{K}$ be a real $n\times
n$ matrix such that its eigenvalues have positive real part,
$\boldsymbol{\Sigma}$ be a real $n\times n$ matrix and
$\boldsymbol{\theta}\in\mathbb{R}^{n}$. Consider the following
OU process%

\begin{equation}
d\boldsymbol{X}(t)=-\boldsymbol{K}\left(  \boldsymbol{X}(t)-\boldsymbol{\theta
}\right)  dt+\boldsymbol{\Sigma}d\boldsymbol{L}(t) , \label{OU}%
\end{equation}
where $\boldsymbol{X}(0)$ is supposed to be independent of $\boldsymbol{L}%
(t)$. Before going into details of our model we discuss the mathematical
framework. The general $n$-dimensional OU process with background driving
L\'{e}vy process (BDLP) is discussed both in \cite{Ba}, where the solution {to \eqref{OU}} is
related to selfdecomposable distributions, and in
\cite{masuda2004multidimensional}. The solution to \eqref{OU} is expressed as
\begin{equation}
\boldsymbol{X}(t)=\left(  I-e^{-t\boldsymbol{K}}\right)  \boldsymbol{\theta
}+e^{-t\boldsymbol{K}}\boldsymbol{X}(\boldsymbol{0})+\int_{0}^{t}%
e^{-(t-u)\boldsymbol{K}}\boldsymbol{\Sigma}d\boldsymbol{L}(u),\quad t\geq0.
\end{equation}

A probability distribution $\eta$ is a stationary distribution for \eqref{OU}
if $\mathcal{L}(\boldsymbol{X}(0))=\eta$ implies that $\mathcal{L}%
(\boldsymbol{X}(t))=\eta$ for every $t$. If the stationary solution exists
then it is unique. It exists a stationary solution if and only if
\begin{equation}
\int_{\mathbb{R}^{n}}\log^{+}|\boldsymbol{x}|\nu_{0}(d\boldsymbol{x})<\infty,
\label{c0}%
\end{equation}
where $\nu_{0}$ is the L\'{e}vy measure of $\boldsymbol{L}(t)$. An equivalent
condition is%
\begin{equation}
\label{c1}\int_{\mathbb{R}^{n}}\log^{+}|\boldsymbol{x}|\eta_{0}%
(d\boldsymbol{x})<\infty,
\end{equation}
where $\eta_{0}=\mathcal{L}(\boldsymbol{L}(1))$.

The stationary solution is $\boldsymbol{K}$-selfdecomposable (see \cite{Ba}).
Furthermore, given a measure $\eta$ on $\mathbb{R}^{n}$ it exists a L\'{e}vy
process such that $\eta$ is the unique stationary solution of \eqref{OU} iff
$\eta$ is $\boldsymbol{K}$-selfdecomposable. Therefore, there is a one to one
relationship between the $\boldsymbol{K}$-selfdecomposable measures $\eta$ on
$\mathbb{R}^{n}$ and the measures $\eta_{0}$ satisfying \eqref{c1}.

Usually the model specification of a OU process with BDLP is performed by
giving the stationary $\boldsymbol{K}$-selfdecomposable distribution $\eta$.
The law of $\eta$ is chosen to belong to a family of selfdecomposable
distributions and $\eta_{0}$ is derived. An important example in the financial
framework of model specification for the one-dimensional case is given in
\cite{BNS}, where the stationary distribution is specified to be a normal
inverse Gaussian (NIG) distribution - which is self-decomposable. The
corresponding process is called NIG-OU process. The multivariate extension of
NIG-OU process is considered in \cite{masuda2004multidimensional}, where the
stationary distribution is chosen to be a multivariate NIG process. Since
\cite{takano1989mixtures} proved that a multivariate normal distribution
subordinated by a generalized gamma distribution is self-decomposable iff the
multivariate normal distribution has zero drift, this assumption is made in
\cite{masuda2004multidimensional}. This restriction leads to symmetric
stationary distributions. {Because of that, and in order to
specify }the dependence structure of the driving process, we choose a different
approach. Instead of specifying the stationary distribution we go the other
way round and construct the model by specifying the BDLP $\boldsymbol{L}(t)$.
{As a result}, the only constraint we have on the choice of $\boldsymbol{L}(t)$
is given by \eqref{c0}.



\subsection{Model specification}

Let now consider the OU process $\boldsymbol{X}(t)$ in \eqref{OU} and assume
that $n=3$, $\boldsymbol{K}$ is lower triangular, $\boldsymbol{\Sigma}$ is lower triangular with unit diagonal elements and $\boldsymbol{L}(t)$ is a L\'{e}vy process on
$\mathbb{R}^{3}$ with independent zero-mean components. {In our economic setting, }the three components $X_{1}(t)$,
$X_{2}(t)$ and $X_{3}(t)$ represent the market interest rates, the deposit
rates and the deposit volume, respectively. Since
$\boldsymbol{K}$ is triangular, so it is $e^{-t\boldsymbol{K}}$. Since
$\boldsymbol{\Sigma}$ is lower triangular with unit diagonal elements, we have%

\begin{equation}
\boldsymbol{\Sigma}d\boldsymbol{L}(t)=\left(
\begin{tabular}
[c]{l}%
$dL_{1}(t)$\\
$\sigma_{21}dL_{1}(t)+dL_{2}(t)$\\
$\sigma_{31}dL_{1}(t)+\sigma_{32}dL_{2}(t)+dL_{3}(t)$%
\end{tabular}
\ \right)
\end{equation}

This specification of the transformations $\boldsymbol{K}$ and
$\boldsymbol{\Sigma}$ captures the economic relationship among the
interest rates, the level of deposit rates and the level of NMDs. The latter
two are built to include components which are specific to the bank and its
customers, together with a systemic factor. In particular, $L_{1}(t)$ is
responsible for systemic shocks in the market interest rates which have an
impact on both the implicit options held by the bank and the customers. The
factor $L_{2}(t)$ captures the idiosyncratic component of deposit rates, which
has also an impact on the level of NMDs. Finally $L_{3}(t)$ represents the
idiosyncratic component specific to the level of NMDs only.

We consider the case of $\boldsymbol{L}(t)$ being a pure jump L\'{e}vy
process, to accommodate possible skewness and kurtosis. In addition, the
Gaussian case, which leads to the standard Ornstein-Uhlenbeck process, is used
as a benchmark to assess the performance of our model. The L\'evy driving
process $\boldsymbol{L}(t)$ is specified to have one dimensional independent
normal inverse Gaussian margins (NIG) -- the univariate NIG process is recalled
in Appendix \ref{NIG}. We choose the NIG specification because it has a good
fit on financial data, it is parsimonious in the number of parameters and with
this choice $\boldsymbol{L}(t)$ satisfies \eqref{c0}.

The process $\boldsymbol{Y}(t)=\boldsymbol{\Sigma}\boldsymbol{L}(t)$ is a
triangular linear transformation of $\boldsymbol{L}(t)$. As a consequence, it
is still a L\'evy process, but its marginal one-dimensional distrbutions at time $t=1$
do not belong in general to the NIG family. {In particular}, ${Y} _{1}(1)$ is NIG distributed,
while the distributions of ${Y} _{2}(1)$ and ${Y}_{3}(1)$ are convolutions of
NIG distributions. The constraint for closure under convolution of a NIG distribution (see
\eqref{constr}) are { in general} not compatible with our model. Indeed, {enforcing convolution conditions means that} we should assume
that $L_{1}(t)$, $L_{2}(t)$ and $L_{3}(t)$ have the same asymmetry and tail
parameters. {  Assuming that those parameters should be equal for interest rates, deposit rates and deposit volumes seems to be unrealistic when we apply the model to real data, as we can see in Section
	\ref{sect:fb} from the analysis of the empirical distributions of residuals (see Figure
	\ref{fig-histograms-EV}).}

\section{Estimation of the model \label{sect:appl}}

We observe the multivariate process $\boldsymbol{X}(t)$ at fixed times
$\ 0=t_{0}<t_{1}<\ldots<t_{n}=T,$ with $\Delta=t_{k+1}-t_{k}$ is a constant.
We have%
\begin{equation}
\boldsymbol{X}(t_{k+1})=\left(  I-e^{-\boldsymbol{K}\Delta}\right)
\boldsymbol{\theta}+e^{-\Delta\boldsymbol{K}}\boldsymbol{X}(t_{k}%
)+\boldsymbol{u}\left(  t_{k}\right)  ,\label{ou-dt}%
\end{equation}
with%
\begin{equation}
\boldsymbol{u}\left(  t_{k}\right)  =\int_{t_{k}}^{t_{k+1}}e^{-(t_{k+1}%
-u)\boldsymbol{K}}\boldsymbol{\Sigma}d\boldsymbol{L}(u).\label{epsilon}%
\end{equation}
The solution (\ref{ou-dt}) is a vector autoregressive process of order one
$VAR\left(  1\right)  $. Letting
\begin{equation}
\boldsymbol{a}=\left(  I-e^{-\boldsymbol{K}\Delta}\right)  \boldsymbol{\theta
,}\quad\boldsymbol{B}=e^{-\boldsymbol{K}\Delta},\label{driftParams}%
\end{equation}
consider%
\begin{equation}
\boldsymbol{X}(t_{k+1})=\boldsymbol{a}+\boldsymbol{BX}(\boldsymbol{t}%
_{k})+\boldsymbol{u}\left(  t_{k}\right)  .\label{VAR1}%
\end{equation}
If $\boldsymbol{L}(t)$ is a multivariate Wiener process, then the conditional
distribution of $\boldsymbol{X}(t_{k+1})|\boldsymbol{X}(t_{k})\,$is normal
with mean $\left(  I-e^{-\boldsymbol{K}\Delta}\right)  \boldsymbol{\theta
}+e^{-\Delta\boldsymbol{K}}\boldsymbol{X}(t_{k})$ and covariance matrix given
by
\[
\text{vec}(\boldsymbol{\Sigma}_{u})=(\boldsymbol{K}\oplus\boldsymbol{K}%
)^{-1}[\boldsymbol{I}-\exp(-(\boldsymbol{K}\oplus\boldsymbol{K})\Delta
)]\text{vec}(\boldsymbol{\Sigma\Sigma}^{\prime})
\]
where vec denotes the stack operator and $\oplus$ denote the Kronecker sum,
since $\boldsymbol{u}\left(  t_{k}\right)  \sim N(0,\boldsymbol{\Sigma}_{u})$
(see \cite{meucci2009review}). Then we can retrieve the original model
parameters from (\ref{driftParams})%
\[
\boldsymbol{K=-}\frac{\log\left(  \boldsymbol{B}\right)  }{\Delta}%
,\quad\boldsymbol{\theta=}\left(  I-e^{-\boldsymbol{K}\Delta}\right)
^{-1}\boldsymbol{a.}%
\]
In the NIG case, the use of ML estimation requires the
inversion of the multivariate characteristic function, which is challenging
from a numerical point of view. From (\ref{ou-dt}) and (\ref{epsilon}), the
error term $\boldsymbol{u}\left(  t_{k}\right)  $ has a triangular structure%
\begin{equation}
\boldsymbol{u}\left(  t_{k}\right)  =\left(
\begin{tabular}
[c]{l}%
$u_{11}(t_{k})$\\
$u_{21}(t_{k})+u_{22}(t_{k})$\\
$u_{31}(t_{k})+u_{32}(t_{k})+u_{33}(t_{k})$%
\end{tabular}
\ \ \ \ \ \ \ \ \ \ \ \right)  \ ,
\end{equation}
where
\[
u_{ij}(t_{k})=\int_{t_{k}}^{t_{k+1}}C_{ij}^{k}\left(  u\right)  dL_{j}(u),
\]
and $C_{ij}^{k}\left(  u\right)  $ is a function of $\boldsymbol{K}$ and
$\boldsymbol{\Sigma}$. If $\Delta$ is small enough, we can assume that
$u_{ij}(t_{k})$ has a NIG distribution with zero mean and variance provided by
Ito's isometry. We can write%
\[
Var\left(  u_{ij}(t_{k})\right)  =s_{ij}^{2}Var\left(  L_{j}\left(  1\right)
\right)  ,\quad i\geq j,
\]
where $L_{j}\left(  1\right)  $ has zero mean and variance $\sigma_{j}^{2}$.
Therefore we estimate the process%
\begin{equation}
\boldsymbol{X}(t_{k+1})=\boldsymbol{a}+\boldsymbol{B}\boldsymbol{X}%
(t_{k})+\boldsymbol{S\varepsilon}(t_{k}),\label{system0}%
\end{equation}
where $S$ is a lower triangular matrix with unit diagonal elements%
\begin{equation}
\boldsymbol{S\varepsilon}(t_{k})=\left(
\begin{tabular}
[c]{l}%
$\varepsilon_{1}(t_{k})$\\
$s_{21}\varepsilon_{1}(t_{k})+\varepsilon_{2}\left(  t_{k}\right)  $\\
$s_{31}\varepsilon_{1}(t_{k})+s_{32}\varepsilon_{2}\left(  t_{k}\right)
+\varepsilon_{3}(t_{k})$%
\end{tabular}
\ \ \ \ \ \ \ \ \ \ \ \ \ \ \ \right)  ,\label{system1}%
\end{equation}
and $\varepsilon_{i}(t_{k})$ are independent identically distributed
$L_{i}\left(  1\right)  \sim\text{NIG}\left(  \alpha_{i},\beta_{i},\delta
_{i},\mu_{i}\right)  ,$ with zero-mean and variance $\sigma_{i}^{2}\,$\ for
all $k.$

The model can be estimated by maximum likelihood (see
\cite{lanne2017identification}). The log-likelihood function of the sequence
$\boldsymbol{\varepsilon}\left(  t_{k}\right)  =\boldsymbol{S}^{-1}\left(
\boldsymbol{X}_{k+1}-\boldsymbol{a}-\boldsymbol{BX}_{k}\right)  ,$
$k=0\ldots,n-1,$ is given by%
\begin{equation}
L\left(  \boldsymbol{\theta}\right)  =\sum_{k=0}^{n-1}\sum_{i=1}^{3}\log
f_{i}\left(  \left(  \boldsymbol{e}^{i}\right)  ^{\prime}\boldsymbol{S}%
^{-1}\left(  \boldsymbol{X}_{k+1}-\boldsymbol{a}-\boldsymbol{BX}_{k}\right)
\right)  , \label{LL}%
\end{equation}
where $\boldsymbol{e}^{i}$ is the $i-th$ unit vector and $f_{i}$ denotes the
log-likelihood function of the sequence of noises $\varepsilon_{i}(t_{k}).$ In
the Gaussian case, $\varepsilon_{i}(t_{k})\sim N\left(  0,\sigma_{i}%
^{2}\right)  ,$ while in the NIG case, we have that $\varepsilon_{i}%
(t_{k})\sim\text{NIG}\left(  \alpha_{i},\beta_{i},\delta_{i},\mu_{i}\right)
,$ $k=0\ldots,n-1$ and $f_{i}$ is given by%
\begin{equation}
f_{i}\left(  x;\alpha_{i},\beta_{i},\delta_{i},\mu_{i}\right)  =n\log\left(
\frac{\alpha_{i}}{\pi}\right)  +n\delta_{i}\sqrt{\alpha_{i}^{2}-\beta_{i}^{2}%
}+\sum_{k=0}^{n-1}\left[  \beta_{i}\delta_{i}\tau_{i}\left(  x\right)  -\log
c_{i,k}+\log H_{1}\left(  \alpha_{i}\delta_{i}c_{i,k}\right)  \right]  ,
\label{LLi}%
\end{equation}
being $H_{1}$ the modified Bessel function of the third kind, $\tau_{i}\left(
x\right)  =\frac{x-\mu_{i}}{\delta_{i}}$ and $c_{i,k}=\sqrt{1+\tau_{i,k}^{2}%
},$ for any $i=1,2,3,$ and $k=0\ldots,n-1$.

In the NIG error specification, we need to estimate the parameter set%
\[
\left\{  \boldsymbol{a},\boldsymbol{B},\boldsymbol{S},\boldsymbol{\sigma
,\alpha},\boldsymbol{\beta},\boldsymbol{\delta},\boldsymbol{\mu}\right\}  .
\]
Let $\gamma_{i}=\sqrt{\alpha_{i}^{2}-\beta_{i}^{2}}.$ By assumption,
$\varepsilon_{i}\,$has zero mean and variance $\sigma_{i}^{2}.$ We set%
\begin{equation}
\mu_{i}+\frac{\delta_{i}\beta_{i}}{\gamma_{i}}=0,\quad\frac{\delta_{i}%
\alpha_{i}^{2}}{\gamma_{i}^{3}}=\sigma_{i}^{2}. \label{contraints}%
\end{equation}
In fact, the constraints will leave us with two free NIG\ parameters only, for
each $i$. The NIG parameters that enters the log-likelihood optimization are
$\left(  \log\gamma_{i},\beta_{i}\right)  ,$ to directly enforce the
positivity of $\gamma_{i}$ and the constraints in (\ref{contraints}). Summing
up, the parameter set to be estimated is%

\[
\boldsymbol{\theta=}\left\{  \boldsymbol{a},\boldsymbol{B},\boldsymbol{S}%
,\boldsymbol{\sigma,\log}\left(  \boldsymbol{\gamma}\right)
,\boldsymbol{\beta}\right\}  ,
\]
while $\boldsymbol{\alpha},\boldsymbol{\delta}$ and $\boldsymbol{\mu}$ are
derived as
\[
\alpha_{i}=\sqrt{\gamma_{i}^{2}+\beta_{i}^{2}},\quad\delta_{i}=\frac
{\sigma_{i}^{2}\gamma_{i}^{3}}{\alpha_{i}^{2}},\quad\mu_{i}=-\frac{\delta
_{i}\beta_{i}}{\gamma_{i}},\quad i=1,2,3.
\]

ML estimators can be computationally demanding in a multidimensional setting.
Initial conditions are derived by a two-step procedure, generalizing the
approach in \cite{chevallier2017} to a multidimensional setting.

Firstly, we
estimate the subset of parameters $\{\boldsymbol{a},\boldsymbol{B}%
,\boldsymbol{S},\boldsymbol{\sigma}\}$ using the least squares approach.
In particular, in the first step, we estimate $\boldsymbol{a}$
and $\boldsymbol{B}$ by multivariate LS estimation of (\ref{VAR1}), which is
equivalent to OLS estimation of the { three} equations separately. The white noise
covariance matrix estimator of $\boldsymbol{\Sigma}_{u}$ of the error term
$\boldsymbol{u}$ is computed from the LS residuals. From
$\boldsymbol{u=S\varepsilon,}$ the covariance matrix $\boldsymbol{\Sigma}_{u}$
is equal to $\boldsymbol{SDS}^{\prime},$ being $\boldsymbol{D}$ the covariance
matrix of the error term $\boldsymbol{\varepsilon}$ with independent
components. By assumption, $\boldsymbol{D}$ is a diagonal matrix with
diagonal elements $\sigma_{i}^{2},$ $i=1,2,3.$ Given the triangular structure
of the matrix $\boldsymbol{S}$, we can derive $\boldsymbol{S}$ and
$\boldsymbol{\sigma}$ \ by equating component-wise the six distinct elements
of the two matrices $\boldsymbol{\Sigma}_{u}$ and $\boldsymbol{SDS}^{\prime}.$

Secondly, we estimate the remaining subset of parameters $\{\boldsymbol{\log
}\left(  \boldsymbol{\gamma}\right)  ,\boldsymbol{\beta}\}$ related to the
error terms $\boldsymbol{\varepsilon}\left(  t_{k}\right)  $ using a ML
approach, keeping $\{\boldsymbol{a},\boldsymbol{B},\boldsymbol{S}%
,\boldsymbol{\sigma}\}$ fixed.

The parameters estimated via the two-step procedure are then used as initial points in the numerical ML procedure.

\section{Modelling non-maturing deposits\label{sect:fb}}

In the present Section, we estimate the model on real data and discuss the fit
of the model, the role of its single components and how those components
interact. The proposed case study is performed using publicly available
aggregate data. Consequently the results obtained are discussed with the only
purpose of exhibiting the main functioning and features of the model. {In fact, it should be stressed that}
the estimates obtained are meaningful only where data relevant for a specific
bank are used to calibrate the model. {Indeed}, the use of less aggregate data enhances
the ability of the model to capture specific features of types of customers or
types of products, reflecting particular market situations such as the
relative positioning of a specific bank within the competitive environment.
Therefore, it would be advisable to have at least a segmentation of the bank'
customers into retail and wholesale categories (\cite{BCBS2016IRRBB}), but the
bank is free to have further refinements in its classification of customers
for internal purposes. In addition, the bank can further discriminate among
types of deposits, identifying e.g. transactional accounts or non-interest
bearing deposits.

Once the scope of the model has been defined and the model has been calibrated
to the relevant data, it will be able to provide concrete answers to the
questions faced by banks in their asset and liability management (ALM).

\subsection{Market rates, deposit rates and volume dynamics}

\label{VEV}

As discussed in the previous section, the three components $X_{1}(t)$,
$X_{2}(t)$ and $X_{3}(t)$ represent the interest rate, the level of deposit
rates and the level of NMDs, respectively. Some further specifications are
needed. The dynamics of interest rates is modelled by the
component $X_{1}(t)$ itself, since negative interest rates are possible and
observed. This means that the dynamics of interest rate follows a traditional
Vasicek model \cite{}. In recent years, deposit rates, represented by the second component
of the model, have turned negative on corporate deposits of several banks of
the Euro area (\cite{altavilla2021there}). Therefore, $X_{2}(t)$ could directly represent
their dynamics. Nevertheless, retail deposit rates are usually positive and
floored at zero. Those floored deposits are hugely relevant for liquidity risk management. In order to encompass the modelling of floored deposits within our setting, we set out an alternative specification where
$X_{2}(t)$ is the deposit log-rates. This means to adopt a dynamics for
deposit rates which is of exponential Vasicek type (see \cite{brigo2007interest}).
On the contrary, {deposit} volumes can never be negative and
cannot be modelled by $X_{3}(t)$ directly. Thus, $X_{3}(t)$ is the dynamics of
log-volumes.

\subsection{Dataset}

\begin{figure}[h]
\centering
\includegraphics[width=0.7\linewidth]{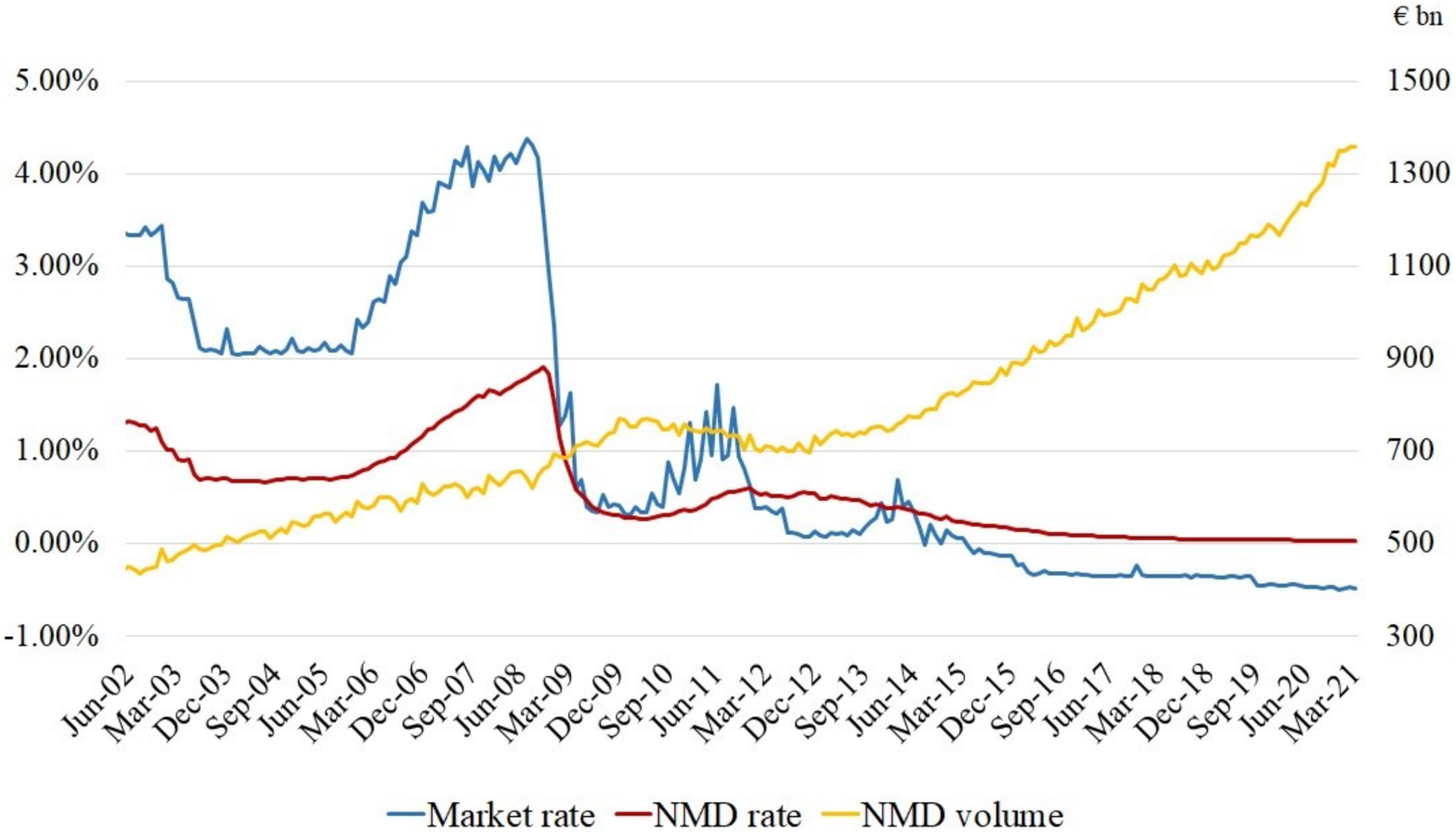}\caption{Time series of market
rates, deposit rates and deposit volume used for the calibration of the
model.}%
\label{fig:dataset}%
\end{figure}

For illustrative purposes three different types of monthly data for market rates,
deposit rates and deposit volumes are used in what follows. Data for EONIA rates are used as a proxy for market interest rates, and are collected from
Bloomberg. The Statistical Database of Banca d'Italia provides publicly available time series from the Italian bank system
on a number of topics. We select the total deposit rates\footnote{code BAM\_MIR.M.1300010.MIR5421.3.950.1000.SBI78.EUR.101.997}
and volumes\footnote{code BAM\_BSIB.M.1070001.52000102.9.101.IT.S1O.1000.997}. The data are shown in Figure \ref{fig:dataset}.

Both the data used for deposit rates and NMDs volume have monthly frequency,
starting from January 2002 up to March 2021, for a total of 231 observations.

\subsection{Estimates of the model parameters}

We now estimate the parameters of equation \eqref{system0}, where $X_{1}(t)$
represents the market rate dynamics, $X_{2}(t)$ represents the deposit log-rate
dynamics and $X_{3}(t)$ represents the log-volume dynamics. In the
optimation of the log-likelihood function $\left(  \ref{LL}\right)  $ we
enforce the following sign constraints%
\begin{equation}
b_{21}>0,b_{31}<0,b_{32}>0,s_{21}>0,s_{31}<0,s_{32}>0,
\label{constraints_est}
\end{equation}
given the financial interpretation of the parameters in our application. In particular, deposit volumes are expected to decrease with market rates and increase with deposit rates.

Table \ref{table_G_NIG_est_EV} provides the estimates of the parameters of
equation \eqref{system0} obtained by ML estimation, together with their translation into
the parameters of equation \eqref{ou-dt}.

\begin{table}[htbp]
	\centering
	\caption{Estimated parameters $\boldsymbol{a}$, $\boldsymbol{B}$, $\boldsymbol{S}$, $\boldsymbol{\sigma}$ and corresponding $\boldsymbol{\theta}$ and $\boldsymbol{K}$.}
	\begin{tabular}{r|r|rrr|}
		\multicolumn{1}{r}{} & \multicolumn{1}{r}{$\boldsymbol{a}$} & $\boldsymbol{B}$ &       & \multicolumn{1}{r}{} \\
		\cline{2-5}    \multicolumn{1}{c|}{\multirow{3}[2]{*}{Gaussian}} & -0.000039 & 0.988688 &       &  \\
		\multicolumn{1}{c|}{} & -0.114547 & 1.734262 & 0.986268 &  \\
		\multicolumn{1}{c|}{} & 0.047423 & -0.060261 & 0.000000 & 0.996912 \\
		\cline{2-5}    \multicolumn{1}{c|}{\multirow{3}[2]{*}{NIG}} & -0.000112 & 0.996328 &       &  \\
		\multicolumn{1}{c|}{} & -0.074274 & 1.130800 & 0.992096 &  \\
		\multicolumn{1}{c|}{} & 0.062410 & -0.147520 & 0.000000 & 0.995876 \\
		\cline{2-5}    \multicolumn{1}{r}{} & \multicolumn{1}{r}{$\boldsymbol{\theta}$} & $\boldsymbol{K}$ &       & \multicolumn{1}{r}{} \\
		\cline{2-5}    \multicolumn{1}{c|}{\multirow{3}[2]{*}{Gaussian}} & -0.003478 & 0.136515 &       &  \\
		\multicolumn{1}{c|}{} & -8.780658 & -21.075061 & 0.165929 &  \\
		\multicolumn{1}{c|}{} & 15.424397 & 0.728377 & -0.000001 & 0.037114 \\
		\cline{2-5}    \multicolumn{1}{c|}{\multirow{3}[2]{*}{NIG}} & -0.030580 & 0.044145 &       &  \\
		\multicolumn{1}{c|}{} & -13.772694 & -13.648619 & 0.095220 &  \\
		\multicolumn{1}{c|}{} & 16.223972 & 1.777170 & -0.000005 & 0.049591 \\
		\cline{2-5}    \multicolumn{1}{r}{} & \multicolumn{1}{r}{$\boldsymbol{\sigma}$} & $\boldsymbol{S}$ &       & \multicolumn{1}{r}{} \\
		\cline{2-5}    \multicolumn{1}{c|}{\multirow{3}[2]{*}{Gaussian}} & 0.002045 & 1.000000 &       &  \\
		\multicolumn{1}{c|}{} & 0.055157 & 10.072156 & 1.000000 &  \\
		\multicolumn{1}{c|}{} & 0.019052 & -0.000031 & 0.000004 & 1.000000 \\
		\cline{2-5}    \multicolumn{1}{c|}{\multirow{3}[2]{*}{NIG}} & 0.002729 & 1.000000 &       &  \\
		\multicolumn{1}{c|}{} & 0.059975 & 5.859505 & 1.000000 &  \\
		\multicolumn{1}{c|}{} & 0.019063 & -0.000246 & 0.007663 & 1.000000 \\
		\cline{2-5}    \end{tabular}
	\label{table_G_NIG_est_EV}%
\end{table}%

Table \ref{table_NIG_step2_EV} provides the estimated parameters for the NIG
specification. The first row refers to the idiosyncratic component of
the market interest rate noise, the second row refers to the
idiosyncratic component $L_{2}(1)$ of the deposit log-rates noise, and the
third row refers to the idiosyncratic component $L_{3}(1)$ of the
deposit log-volumes. The last two columns of Table \ref{table_NIG_step2_EV}
show the estimated annual skewness and kurtosis of the three scaled
idiosyncratic components representing the marginal idiosyncratic components of
the error. Notice that in our dataset market rates show high negative
skewness and high excess kurtosis that cannot be captured in the Gaussian case.

\begin{table}[ptbh]
\caption{ Estimated parameters of the idiosyncratic components
$\boldsymbol{L}(1)$ for the NIG specification (second step). The last two
columns show the annual skewness and kurtosis.}%
\label{table_NIG_step2_EV}%
\centering%
\begin{tabular}
[c]{lrrrrrr}
& \multicolumn{1}{c}{$\alpha$} & \multicolumn{1}{c}{$\beta$} &
\multicolumn{1}{c}{$\delta$} & \multicolumn{1}{c}{$\mu$} &
\multicolumn{1}{c}{Skewness} & \multicolumn{1}{c}{Kurtosis}\\\hline
$L_{1}(1)$ & 52.52986 & -9.29901 & 0.00037 & 0.00007 & -1.91 & 46.77\\
$L_{2}(1)$ & 17.09158 & -9.14173 & 0.03709 & 0.02348 & -1.10 & 6.00\\
$L_{3}(1)$ & 71.33072 & 12.01585 & 0.02483 & -0.00424 & 0.19 & 3.48\\\hline
\end{tabular}
\end{table}

\begin{figure}[tbh]
\begin{center}
\includegraphics[scale=0.55]{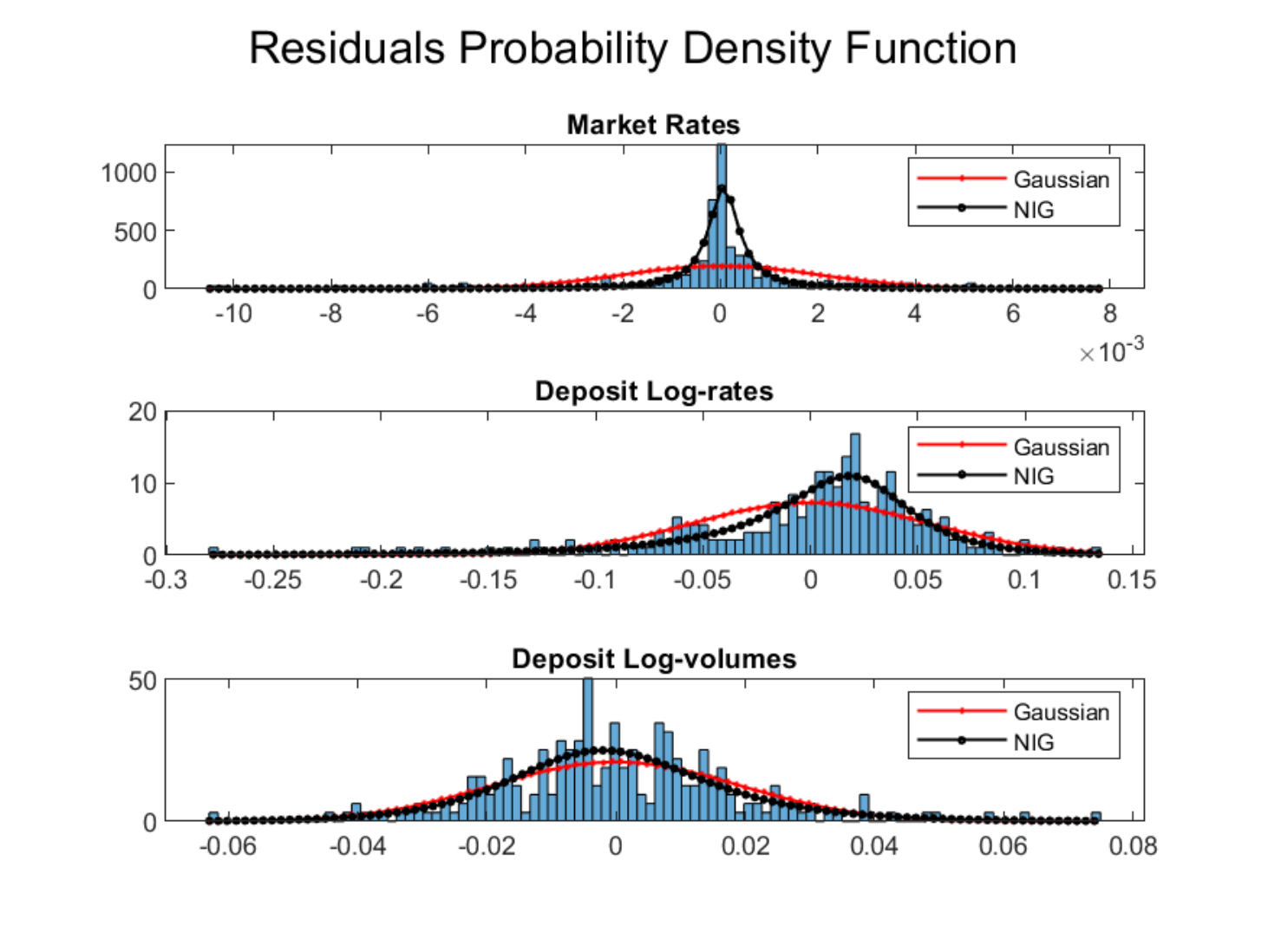}
\end{center}
\caption{Histogram of the empirical residuals and the corresponding Gaussian
and NIG fitted pdf.}%
\label{fig-histograms-EV}%
\end{figure}

Figure \ref{fig-histograms-EV} shows how the NIG assumption critically
improves the fit of the empirical distribution of the idiosyncratic errors.
This is particularly evident for the market rates and deposit log-rates.

\section{Application: liquidity risk management}

\label{sec:lrm}

In this section we make use of our estimated multivariate L\'{e}vy-driven OU
process as a tool for liquidity risk management. We recall that the
three components of the process model interest rates $X_{1}(t)$, deposit
log-rates $X_{2}(t)$ and log-volumes $X_{3}(t)$ of NMDs.

According to \cite{BCBS2016Liquidity}, banks should have a sound process for
identifying, measuring, monitoring and controlling liquidity risk, including a
robust framework for projecting cash flows arising from assets, liabilities
and off-balance sheet items over appropriate time horizons. It is important,
in particular, to correctly assess the
“stickiness” of funding sources, i.e.
the tendency of these sources not to run off quickly under stress. Among the
factors that influence the
“stickiness” of deposits, the Basel
standards cite the interest-rate sensitivity, the size of the deposit, the
geographical location of depositors and the deposit channel. The first of
these factors is explicitly modelled in our framework, while the others, as
previously mentioned, can be captured by means of calibration to historical
data for specific types of deposits or depositors.

\begin{figure}[h]
\begin{center}
\includegraphics[scale=1]{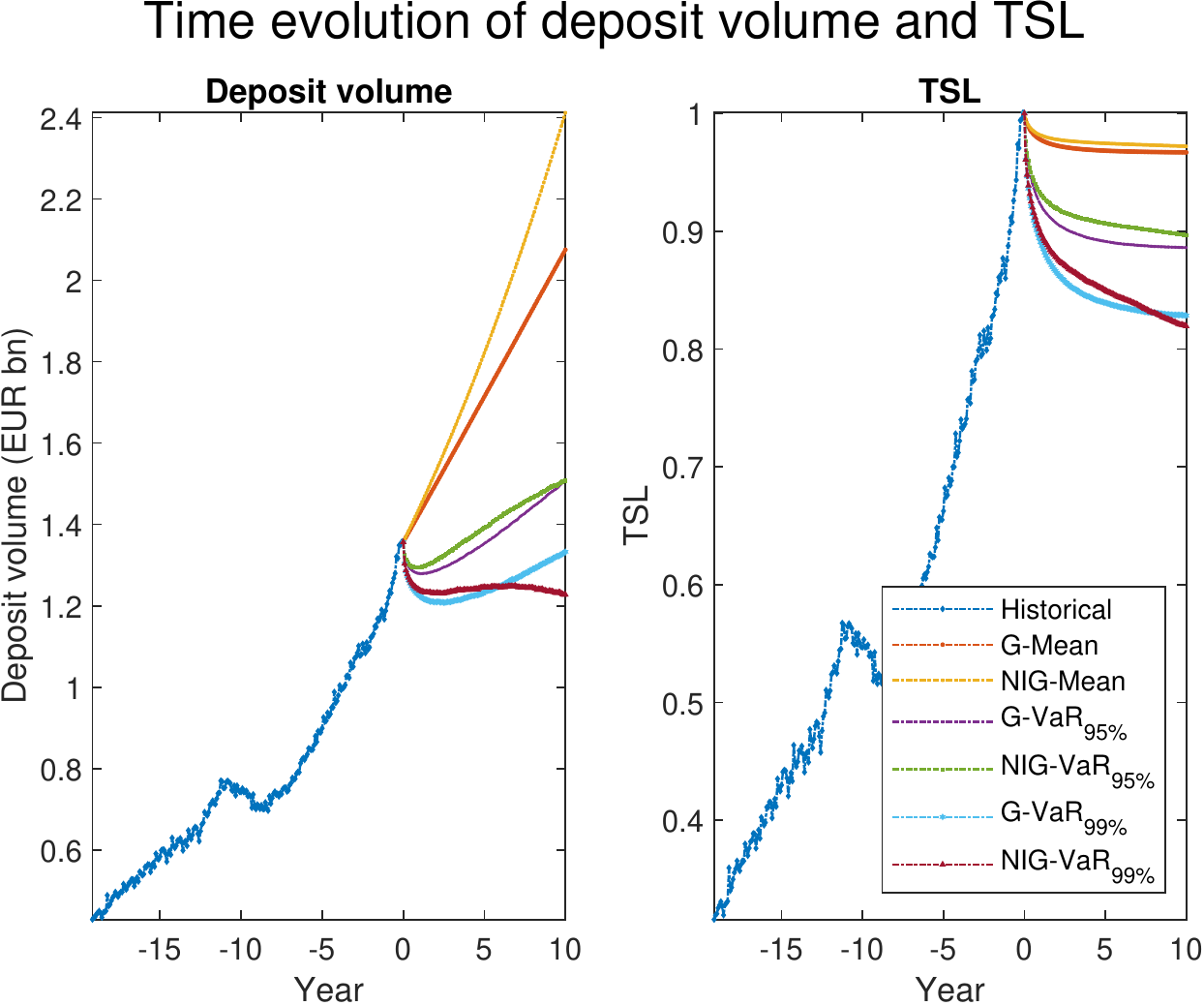}
\end{center}
\caption{Time evolution of the deposit volume (EUR bn) and $TSL$ at
different confidence levels for the Gaussian and NIG specifications. The
initial deposit volume is equal to 1,356 EUR bn. }%
\label{fig_base_case}%
\end{figure}

We proceed by Monte Carlo simulation, generating 100,000 paths of the joint
dynamics of the risk factors on a time grid with step size of one month and
time horizon up to 10 years. We focus on the lowest quantiles of the projected
deposit volume distribution. We define the Value-at-Risk at time $t\,\ $of the
deposit volumes at the confidence level $\alpha\,\,$as the quantile of order
$1-\alpha\,\,$ of the distribution of deposit volume at any given time $t$%
\[
VaR_{\alpha}\left(  D\left(  t\right)  \right)  =\inf\left\{  d\in\left(
0,+\infty\right)  :\mathbb{P}\left(  D\left(  t\right)  \leq d\right)
\geq1-\alpha\right\}  .
\]
The left-hand side of Figure
\ref{fig_base_case} shows the historical evolution
of the deposit volumes in the past 19 years, corresponding to our sample
period, and their projected evolution over the next 10 years. The projected
evolution is summarized by its expected value, value-at-risk levels and
expected shortfall. Results are presented both for the NIG specification of
the BDLP and the Gaussian one, with the latter serving as a benchmark for the
first. From the plot, one can appreciate differences and similarities of the
two specifications. While similar results are obtained in terms of the
$VaR_{95\%}$ of the evolution of deposits, differences can be observed in
terms of $VaR_{99\%}$. The phenomenon is a clear consequence
of the \textquotedblleft fat tails\textquotedblright\ embedded in the NIG distribution.

However in many practical cases, banks prefer to look at different metrics,
which could be proven to be more meaningful than the distribution at time $t$
of the NMDs. Indeed, very often banks are interested in identifying the
minimum level of deposit volumes up to any given time $t$, with a given
confidence level. This would be the actual amount of funds available for
reinvestment on the horizon $\left[  0,t\right]  $. We define the lowest level
of deposit volumes up to time $t$ as%
\[
M\left(  t\right)  =\min_{0\leq s\leq t}D\left(  s\right)  .
\]
We normalize the quantiles of $M\left(  t\right)  $ by the initial value of
the deposit volume (see e.g. \cite{kalkbrener2004risk} and \cite{castagna2017}%
). We obtain the so-called Term Structure of Liquidity with confidence level
$\alpha$, used in the following as reference metrics%
\[
TSL_{\alpha}\left(  t\right)  =\frac{VaR_{\alpha}\left(  M\left(  t\right)
\right)  }{D\left(  0\right)  },\,t\geq0.
\]
By construction, $TSL_{\alpha}\left(  t\right)  $ is a non-increasing function
of $t\,$. The normalization with respect to the initial value ${D\left(
0\right)  }$ eases the comparison among different clusters of depositors when
a financial institution implements the model to internal data. Table
\ref{table_tsl} shows $TSL_{\alpha}\left(  t\right)  $ for our selected grid
of confidence levels and time horizons, while the right-hand side of Figure
\ref{fig_base_case} displays the projected evolution of the $TSL_{\alpha}$.
From the projection of the $TSL_{\alpha}$, one can appreciate the stability of
the deposit volume over time. In our illustrative example, $90\%$ and $82\%$ of the NMDs
are expected to be available after 10 years with $95\%$ and $99\%$ confidence level, respectively
($89\%$ and $83\%$ in the Gaussian specification of the BDLP, similarly to the NIG specification), proving to be very stable. Analysing the expected shortfall at $97.5\%$, the results are slightly more pronounced for the NIG specification ($81\%$) than for the Gaussian one ($83\%$), signalling the ability of the NIG specification to capture a larger part of the risks embedded in the historical data.

\subsection{Stressed parameters and inclusion of bank runs}

\cite{BCBS2016Liquidity} stresses the importance of the assumptions used in
projecting future cash flows. These assumptions should be adjusted
according to market conditions or bank-specific circumstances, therefore
they should include all the events that could entail a significant risk in terms of
liquidity management for a bank. The multivariate L\'{e}vy-driven OU model
presented in this paper allows for the inclusion of bank run events, i.e.
events occurring when a large number of customers withdraw their deposits in a
relatively short period of time. These events can be triggered by a number of
different causes, but as common characteristics they are typically
unpredictable and idiosyncratic to one or few banks. They can have a severe
impact on a bank, as the more people withdraw their deposits, the more the
stability of a bank is undermined, potentially encouraging further
withdrawals. Although our model is well-suited for capturing rare but severe
events, these events are quite infrequent and it is difficult to observe them in
the time series used for the calibration of the model. Therefore, we propose
here an alternative strategy for capturing such events within our model,
outside from the calibration to historical data. As a starting point of our
strategy, we take a real case of bank run. In 2019, Metro Bank PLC, one of the
major banks in UK, experienced a severe outflow of deposits in a relative
short period of time. In the \textquotedblleft Half Year 2019
Results\textquotedblright,\footnote{https:$//$www.metrobankonline.co.uk$/$%
globalassets$/$documents$/$investor$\_$documents$/$%
trading-announcement-h1-2019.pdf} the bank reported a $25\%\,$ volume outflow
of deposits from commercial customers in the period from 31 December 2018 to
30 June 2019. According to what reported by the bank, the total deposits'
decrease ($-13\%$) in that reported 6-month period was \textquotedblleft%
\textit{driven by a limited number of commercial customers withdrawing
deposits during intense speculation in February and May}\textquotedblright. As mentioned above, on
the one hand, that event can be incorporated in the model via calibration, but that is possible only if the model is calibrated to the specific Metro Bank PLC historical data, given that that event has been recorded in those historical data only. On the other hand, an
alternative strategy can be to calibrate the model to the available data and then to stress the calibrated parameters in order to
enable the model to reproduce that stress event with a predetermined confidence
level. Several approaches are possible and what we propose here is one of the
many. Firstly, we define the relative amount of deposit volume outflow with confidence
level $\alpha$ at time $t+h$, given the deposits' level at time $t$,

\[
RDO_{\alpha}\left(t,h \right)  = 1-VaR_{\alpha}\left(   \frac{D\left(
	t+h\right)}{D\left(  t\right)  } \vert \mathcal{F}_{t}\right)
,
\]
where $\mathcal{F}_{t}\,$is the natural filtration of process $\boldsymbol{X}%
(t)$ in \eqref{system0}.
Secondly, given that we want the stress event (i.e. $25\%$ deposit outflow in 6 months) to be reproduced by the model from $t$ to $T$ (the time horizon of our simulation), we need to calibrate the parameters of the model so to reproduce, at least on average, that effect from $t$ to $T$.
Considering that we observe the multivariate process $\boldsymbol{X}(t)$ at fixed times
$\ t=t_{0}<t_{1}<\ldots<t_{n}=T$, with $\Delta=t_{k+1}-t_{k}$ constant, and that for each $t_{k}$ we can compute the relative amount of deposit outflow with confidence level $\alpha$ and time horizon $t_{k}+h$, where $h > \Delta$, the following measure is defined

\[
\overline{RDO}_{\alpha}\left(t,h \right)  = \frac{1}{m+1}\sum_{k=0}^{m} RDO_{\alpha}\left(t_k,h \right)
,
\]
where $m=n-\frac{h}{\Delta}$. Thirdly, the parameters of the model are
\textquotedblleft stressed\textquotedblright\ (i.e. partially re-calibrated)
in order to enable the model to produce on average the stress event (i.e. an deposit volumes outflow
of at least $25\%$) at any time $t_{k}+h$, given the level of
NMDs at time $t_{k}$, with a predetermined level of probability $1-\alpha$
(arbitrarily set at $0.1\%$). Given that the event to be captured by the model is
idiosyncratic and affects the level of NMDs only, we focus on the parameters
governing the probability distribution of the noise $\varepsilon_{3}$ in the
dynamics of the deposit volume $X_{3}(t)$ in \eqref{system0}.
In order to do so, we look for the combination
of the NIG parameters of the deposit volumes idiosyncratic component which results in an $\overline{RDO}_{99.9\%}\left(t,\textrm{6 months}\right)=25\%$, where $\Delta=$ 1 month and $T =\, t\; + $ 10 years. Considering that the event
we want to reproduce belongs to the negative tail of the distribution of the deposit volumes, when we iteratively look for the set of \textquotedblleft stressed\textquotedblright\ parameters, the mean and variance of the
error term are kept at the level calibrated to historical data, while both
the skewness and kurtosis are allowed to change.

\begin{figure}[h]
\centering
\includegraphics[width=0.7\linewidth]{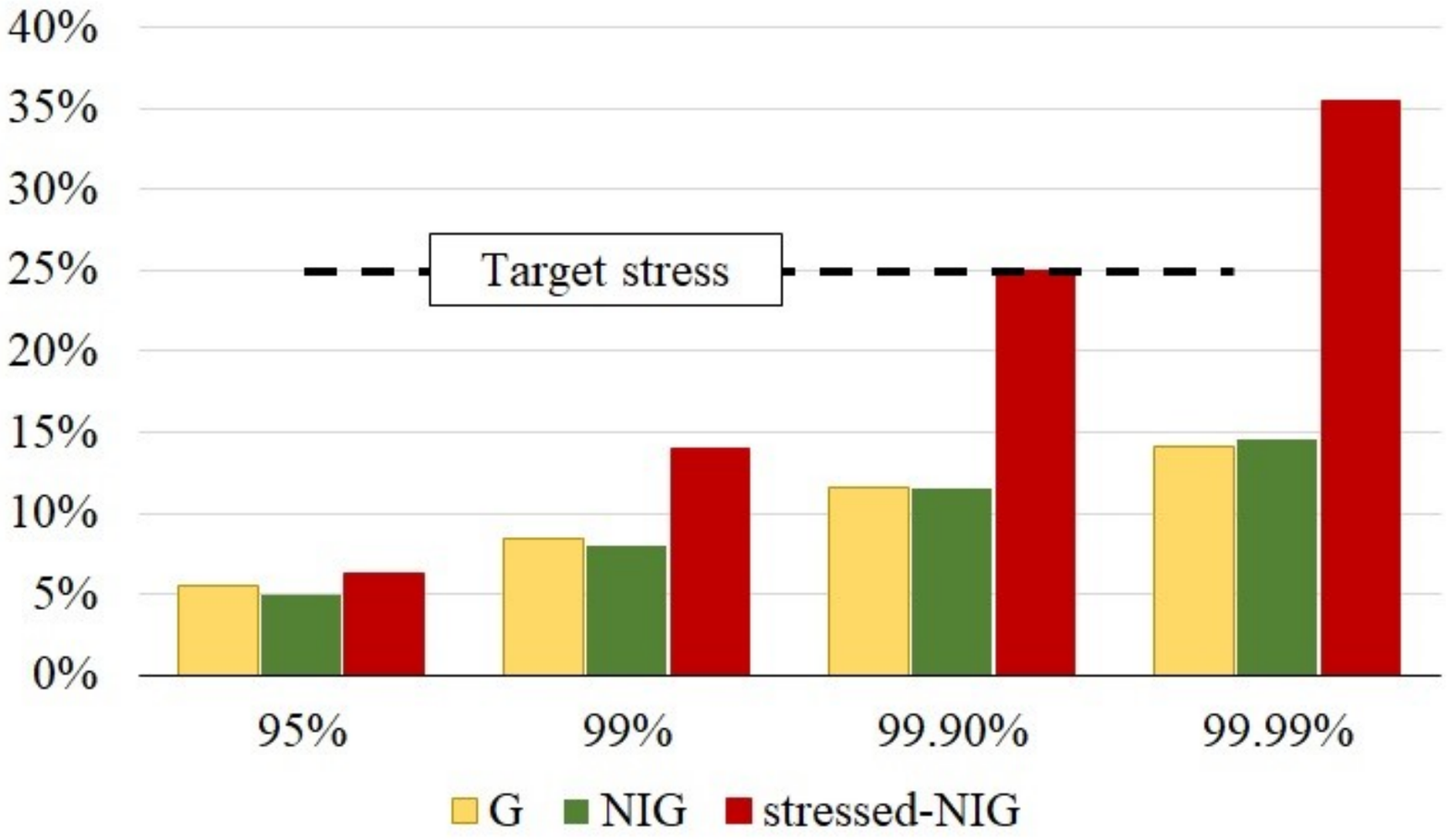} \caption{Calibration of
$\overline{RDO}_{\alpha}\left(t,\textrm{6 months}\right)$ to a target stress event (outflow of
deposit volume of at least $25\%$).}%
\label{fig:stress_RUN6M}%
\end{figure}

\begin{figure}[h]
\begin{center}
\includegraphics[scale=1]{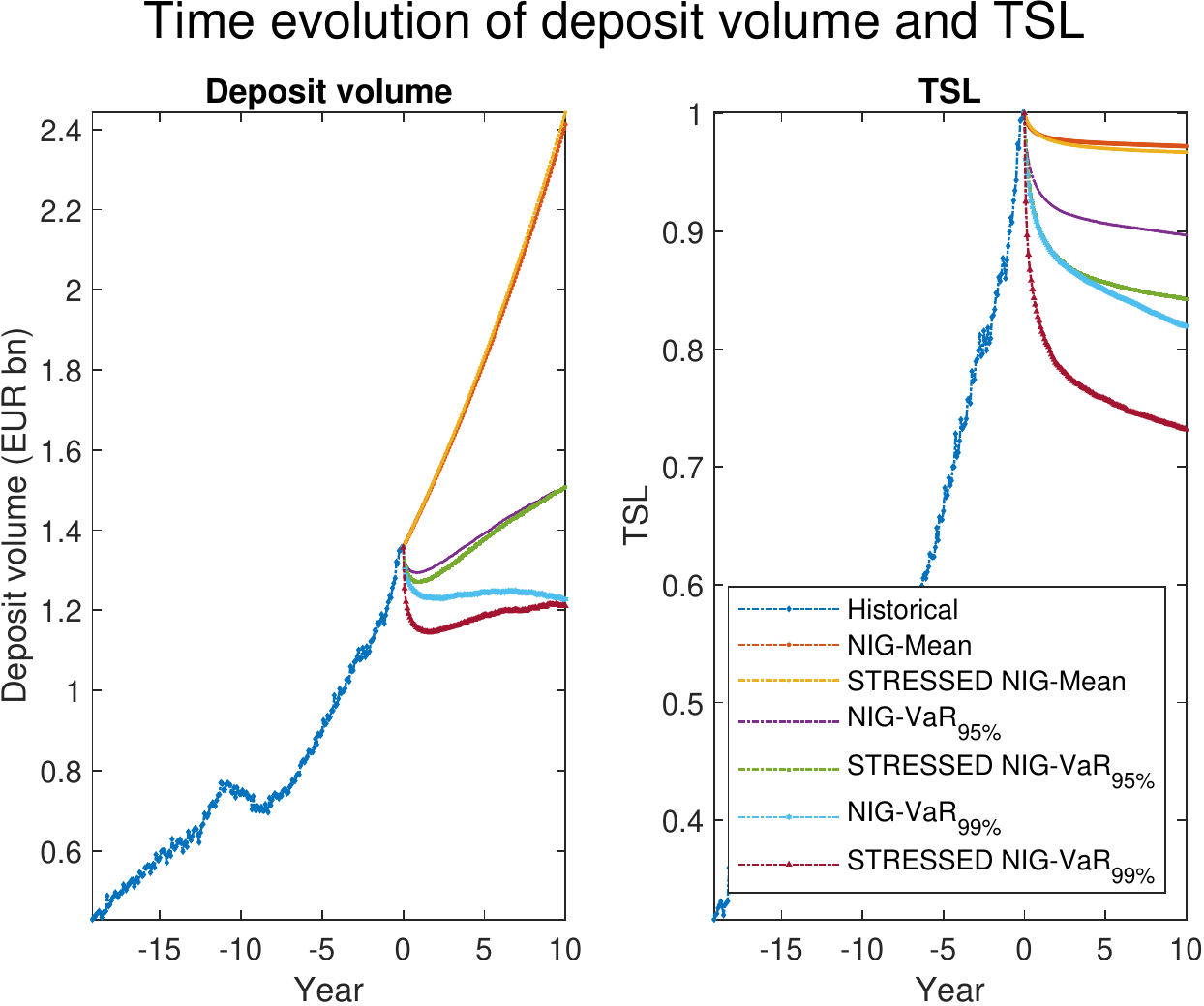}
\end{center}
\caption{Time evolution of the deposit volume (EUR bn) and $TSL$ at
different confidence levels for the NIG and stressed-NIG specification. The
initial deposit volume is equal to 1,356 EUR bn.}%
\label{fig:stressed_case_nig}%
\end{figure}

The outcome of the calibration of the parameters to an $\overline{RDO}_{99.9\%}\left(t,\textrm{6 months}\right)$ $=25\%$ is represented in Figure \ref{fig:stress_RUN6M}. In
particular, one can appreciate that the stressed parameters imply a significant change in
the left tail of the distribution only. The observed average relative amount of deposits outflow in 6 months, i.e.  $\overline{RDO}_{95\%}\left(t,\textrm{6 months}\right)$ represented in the first three columns of Figure \ref{fig:stress_RUN6M}, is mainly driven by the variance of the distribution and consequently
the observed values of that measure are close for the three cases at hand (Gaussian, NIG and stressed NIG). Setting $\overline{RDO}_{99.9\%}\left(t,\textrm{6 months}\right)=25\%$ (see ninth column of Figure \ref{fig:stress_RUN6M}) translates
in an annual skewness of the idiosyncratic component of the deposit volume
equal to $-1.74$ and an annual kurtosis equal to $5.17$. By setting
$\alpha=269.4450$, $\beta=-256.7294$, $\delta=0.0027$, and $\mu=0.0086$, we
obtain the stressed evolution of deposit volume and $TSL\,$, as shown in Table
\ref{table_tsl} and Figure \ref{fig:stressed_case_nig}.

In Figure \ref{fig:stressed_case_nig} one can appreciate the difference
between the calibrated NIG specification and the stressed-NIG, which is
significant for the $VaR_{95\%}$ and even more pronounced for the $VaR_{99\%}$
of both the evolution of deposit volume and the $TSL\,$. As expected, the
inclusion of the possibility of bank run event in the model generates an
increasing of the risk measures. These measures depend on the likelihood of
that event, as well as on its magnitude.

\begin{table}[ptbh]
\caption{ VaR and expected shortfall of the Term Structure of Liquidity at
different time horizons and confidence levels. }%
\label{table_tsl}
\centering
\begin{tabular}
[c]{c|ccc|ccc|ccc}
& \multicolumn{3}{c|}{TSL-Gaussian} & \multicolumn{3}{c|}{TSL-NIG} &
\multicolumn{3}{c}{Stressed TSL-NIG}\\\hline
& Var$_{95\%}$ & Var$_{99\%}$ & ES$_{97.5\%}$ & Var$_{95\%}$ & Var$_{99\%}$ &
ES$_{97.5\%}$ & Var$_{95\%}$ & Var$_{99\%}$ & ES$_{97.5\%}$\\\hline
1YR & 92\% & 89\% & 89\% & 93\% & 90\% & 90\% & 90\% & 82\% & 82\%\\
3YR & 90\% & 85\% & 85\% & 91\% & 87\% & 87\% & 87\% & 77\% & 77\%\\
5YS & 89\% & 84\% & 84\% & 91\% & 85\% & 85\% & 86\% & 76\% & 75\%\\
10YR & 89\% & 83\% & 83\% & 90\% & 82\% & 81\% & 84\% & 73\% & 73\%\\\hline
\end{tabular}
\end{table}

\section{Conclusion and further applications}

\label{conclusion}

The contribution of the paper to the operational research literature relates to
the management of NMDs for banks. We build a multivariate OU process to model
the interactions among market interest rates, deposit rates and deposit
volumes. By specifying the driving process to be a L\'evy process, we are able
to incorporate rare but significant events in the
liquidity risk management of a bank. This paper also clarifies how the proposed model can be estimated via an ML approach.

We also propose an operational procedure to
stress the calibrated parameters in order to enable the model to reproduce
rare but significant events (e.g. bank runs) with a predetermined confidence level. As a starting point of our strategy, we take a real case of bank run. We show that the stressed
parameters produce significantly increased measures of risk. Moreover, our
risk factor model can also be used to perform scenario analysis. Focusing on
scenarios for the market rates variable, the analysis can be based on
internally selected interest rate shock scenarios, historical or hypothetical
interest rate stress scenarios, Basel-prescribed interest rate shock scenarios
or any additional interest rate shock scenarios required by supervisors (see
\cite{BCBS2016IRRBB}).

In addition to applications in the liquidity risk management, our
multivariate L\'{e}vy-driven OU model can be used for the purpose
of interest rate risk (in the Banking Book - IRRBB) management. As underlined
in \cite{BCBS2016IRRBB}, IRRBB is a material risk faced by banks, with this
materiality expected to be more pronounced when interest rates may normalise
from the current low levels. When interest rates change, the economic value of
the NMDs changes, as well as the bank’s earning capacity,
measured by its net interest income (NII). In particular, the model can be
used in the construction of a bond portfolio with fixed maturities,
replicating the price and delta profile of the NMDs. This portfolio can be
used then in the computation of the aggregated economic value and earning
measures.

\section*{Acknowledgement}

Marina Marena and Patrizia Semeraro gratefully acknowledges financial support
from the Italian Ministry of Education, University and Research (MIUR),
"Dipartimenti di Eccellenza" grant 2018-2022. \appendix
\appendixpage

\section{Normal inverse Gaussian process}

\label{NIG}

The univariate NIG process has been defined by \cite{barndorff1995normal}. A
NIG process with parameters $\alpha>0,\,-\alpha<\beta<\alpha,\,\delta>0$ is a
L\'{e}vy process $\{L_{NIG}(t),t\geq0\}$ with characteristic function at time
1
\[
\psi_{NIG}(u)=\exp\left(  -\delta\left(  \sqrt{\alpha^{2}-(\beta+iu)^{2}%
}-\sqrt{\alpha^{2}-\beta^{2}}\right)  \right)  .
\]
The NIG process has been chosen for its ability to accommodate skewness and
kurtosis and for its analytical tractability. We recall below the mean $m$,
the variance $v$, the skewness $s$ and the kurtosis $k$ of the NIG
distribution:
\[
m=\frac{\delta\beta}{\sqrt{\alpha^{2}-\beta^{2}}},
\]

\[
v={\alpha^{2}\delta}{({\alpha^{2}-\beta^{2}})^{-\frac{3}{2}}},
\]

\[
s=3\beta{\alpha^{-1}\delta^{-\frac{1}{2}}}{({\alpha^{2}-\beta^{2}})^{-\frac
{1}{4}}},
\]

\[
k=3(1+\frac{\alpha^{2}+4\beta^{2}}{\delta\alpha^{2}\sqrt{\alpha^{2}-\beta^{2}%
}}).
\]

The NIG class is closed under convolution provided the parameters $\alpha$ and
$\beta$ are fixed, i.e.
\begin{equation}
\label{constr}NIG(\alpha, \beta, \delta_{1})*NIG(\alpha, \beta, \delta
_{2})=NIG(\alpha, \beta, \delta_{1}+ \delta_{2}),
\end{equation}

$\alpha$ and $\beta$ are the skewness and tail parameters.

\newpage
\bibliographystyle{apalike}
\bibliography{biblio}

\end{document}